\documentclass[twocolumn]{aastex62}
\usepackage{makecell, multirow}
\setcellgapes{3pt}

\usepackage{graphicx}% Include figure files
\usepackage{dcolumn}% Align table columns on decimal point
\usepackage{bm}% bold math
\usepackage{color}
\newcommand{\msun}{\ensuremath {\rm M}_{\odot}}
\newcommand{\rsun}{\ensuremath {\rm R}_{\odot}}

\usepackage{array}
\newcolumntype{P}[1]{>{\centering\arraybackslash}p{#1}}

\newcommand\NScount{53.2}
\newcommand\BHcount{19.5}
\newcommand\WDcount{26.2}
\newcommand\STcount{1.2}

\newcommand\EMRI{$(3-8)\times 10^{-5}$}
\newcommand\Plunge{$(1-5)\times 10^{-5}$}
\newcommand\TDE{$2 \times 10^{-5}$}
\newcommand\XrayBin{$(6.3-9.9) \times 10^{-3}$}
\newcommand\GWmerger{$(0.17 - 1.2)\times 10^{-3}$}
\newcommand\BinDisr{$(0.68-3.9)\times 10^{-2}$}

\newcommand\NSemri{$\sim 88$}
\newcommand\BHemri{$\sim 6$}
\newcommand\WDemri{$\sim 6$}

\shortauthors{Hoang et al. }

\begin{document}

\title{Binary Natal Kicks in the Galactic Center: X-ray Binaries, Hypervelocity Stars, and Gravitational Waves}

\correspondingauthor{Bao-Minh Hoang}
\email{bmhoang@astro.ucla.edu}

\author{Bao-Minh Hoang}
\affiliation{Department of Physics and Astronomy, University of California, Los Angeles, CA 90095, USA}
\affiliation{Mani L. Bhaumik Institute for Theoretical Physics, Department of Physics and Astronomy, UCLA, Los Angeles,
CA 90095}
\author{Smadar Naoz}
\affiliation{Department of Physics and Astronomy, University of California, Los Angeles, CA 90095, USA}
\affiliation{Mani L. Bhaumik Institute for Theoretical Physics, Department of Physics and Astronomy, UCLA, Los Angeles,
CA 90095}

\author{Melodie Sloneker}
\affiliation{Department of Physics and Astronomy, University of California, Los Angeles, CA 90095, USA}
\affiliation{Mani L. Bhaumik Institute for Theoretical Physics, Department of Physics and Astronomy, UCLA, Los Angeles,
CA 90095}

\begin{abstract}
Theoretical and observational studies suggest that stellar binaries exist in large numbers in galactic nuclei like our own Galactic Center. Neutron stars (NSs), and debatedly, black holes (BHs) and white dwarfs (WDs), receive natal kicks at birth. In this work we study the effect of two successive natal kicks on a population of stellar binaries orbiting the massive black hole (MBH) in our Galactic Center. These natal kicks can significantly alter the binary orbit in a variety of ways, and also the orbit of the binary around the MBH. We found a variety of dynamical outcomes resulting from these kicks, including a steeper cusp of single NSs relative to the initial binary distribution. Furthermore, hypervelocity star and binary candidates, including hypervelocity X-ray binaries, are a common outcome of natal kicks. In addition, we show that the population of X-ray binaries in the Galactic Center can be used as a diagnostic for the BH natal kick distribution. Finally, we estimate the rate of gravitational wave (GW) events triggered by natal kicks, including binary mergers and EMRIs.
\end{abstract}

\section{Introduction}\label{sec:Intro}

Most galaxies, including our own, host a MBH at its center \citep[e.g.][]{Kormendy+95,Ghez+00,Ghez+08,Kormendy+Ho}. Around these MBHs are the incredibly dense structures composed of stars and stellar remnants, called nuclear star clusters (NSCs). 
 
 Both theoretical and observational evidence suggest that binaries are common in NSCs. On the theoretical side, \citet{Stephan+16} showed that as many as 70\% of the binaries formed in an initial star formation episode survive after a few Myrs. Furthermore, \citet{Rose+20} showed that these surviving young (few Myrs) binaries can have a wide range of orbital configurations and separations.  \citet{Naoz+18} showed that a high binary fraction can explain many puzzling aspects of the young stellar disk in our own Galactic Center \citep[e.g.,][]{Yelda+14}.  

On the observational side, recent works have suggested that binaries are prevalent in our Galaxy for massive stars (about $70\%$ for ABO spectral type stars) \citep[e.g.][]{Raghavan+10,Moe+17}. Thus, we can reasonably infer a comparably large binary fraction for massive stars in NSCs. Furthermore, the large number of observed X-ray sources in the central 1 pc suggest relatively large numbers of binaries containing BHs and NSs in the Galactic Center \citep{Muno+05,Hailey+18}. Moreover, there are three confirmed binaries in the inner $\sim 0.2$~pc \citep[e.g.,][]{Ott+99,Martins+06,Rafelski+07,Pfuhl+14}, with possibly more candidates \citep[e.g.,][]{Jia+19,Gautam+19}.

For massive binaries, a potentially important effect is the ``natal" kick that neutron star progenitors (and debatedly, black hole progenitors), receive from supernova explosions. Observations of pulsar proper motions have shown that neutron stars (NSs) receive large kicks of hundreds of km/s \citep[e.g.][]{Hansen+Phinney97,Lorimer+Bailes97,Cordes+Chernoff98,Fryer+99,Hobbs+04}. Natal kicks have been shown to explain the spin-orbit misalignment of pulsar binaries \citep{Lai+95,Kalogera96,Kaspi+96,Kalogera+98,Kalogera00}. Furthermore, it has been suggested that hypervelocity stars \citep{Zubovas+13,Bortolas+17,Fragione+17,Lu+19}, and EMRIs \citep[e.g.,][]{Lu+19,Bortolas+19}, can result from natal kicks.

Recently, \citet{Lu+19} explored natal kicks in hierachical triple systems, including stellar-mass binaries near an MBH \citep[see also][]{Hamers+18}. They found that the natal kicks in these systems can lead to the inner binary shrinking as well as expanding, leading to many possible observable phenomena. In this work, we explore the dynamical and observable consequences of two successive natal kicks on a population of stellar binaries in the Galactic Center. Below we give a brief summary of the steps we take in our study and the structure of this paper:
\begin{figure*}
    \centering
    \includegraphics[width=\linewidth]{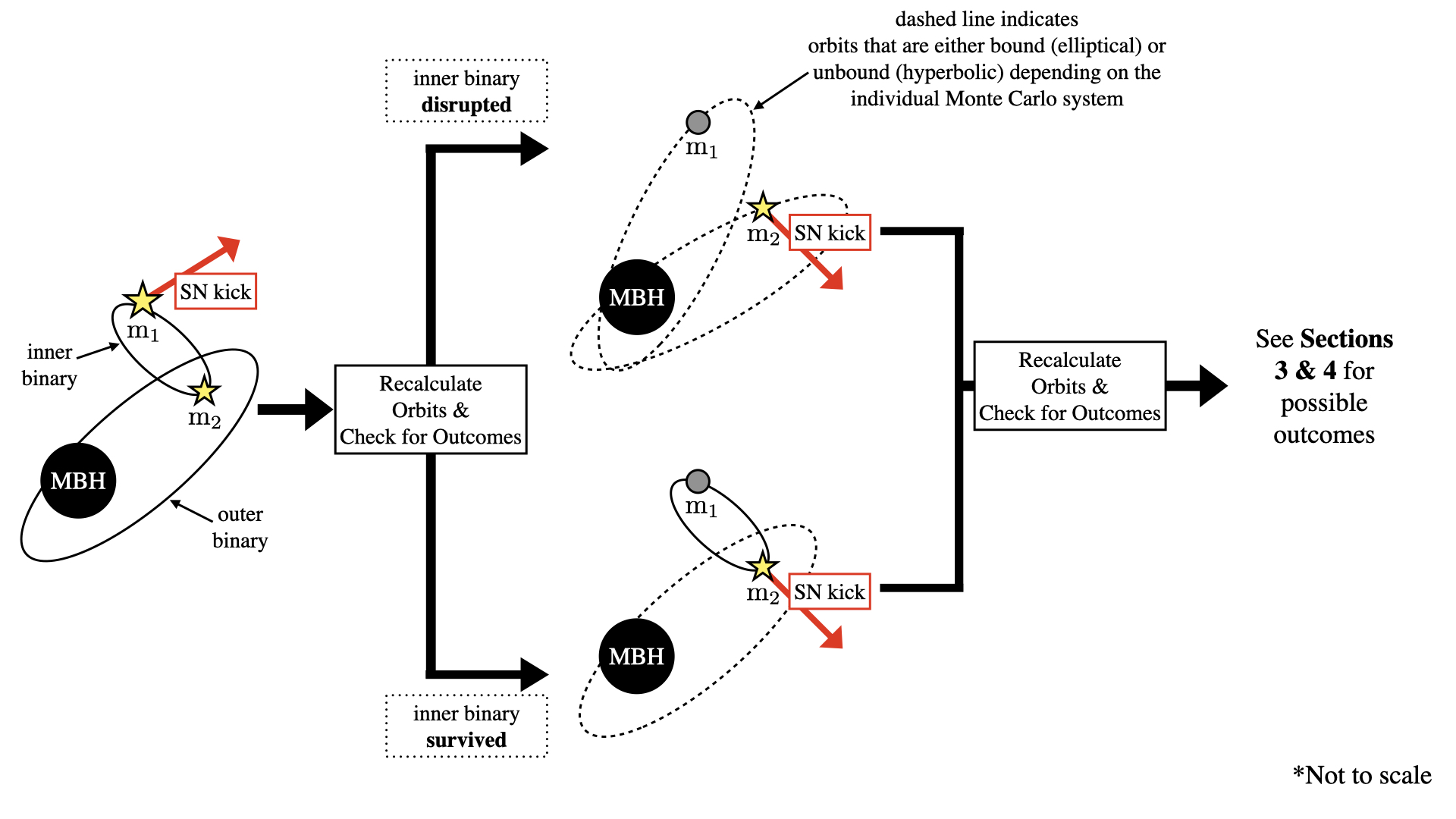}
    \caption{A simplified schematic of our analysis process. Note that this schematic does not illustrate every single possible outcome of natal kicks on a binary.}
    \label{fig:schematic}
\end{figure*}

\begin{enumerate}
    \item We generate a population of massive stellar binaries. This is described in Section \ref{sec:MC}.
    \item We evolve the stellar binaries up to the $m_1$'s natal kick and eliminate any binaries that undergo Roche Lobe overflow before $m_1$'s natal kick. This process is described in Section \ref{sec:preSN}.
    \item We apply the natal kick to $m_1$ and recalculate the orbits. We check the new orbits for any natal kick-induced observable phenomena (e.g. X-ray Binary, hypervelocity star). We then apply the natal kick to $m_2$ and repeat the process of checking for outcomes. We describe how we apply the natal kicks in more detail in Section \ref{sec:SNkicks}. We describe the criteria we used to classify the outcomes and describe the results of our Monte Carlo simulations in Sections \ref{sec:single_outcomes} and \ref{sec:binary_outcomes}.
\end{enumerate}

A simplified schematic of our steps is given in Figure \ref{fig:schematic}, and a summary of the outcomes of our simulations is given in bar chart form in Figure \ref{fig:AggregateStats}.

\section{Methodology}
We employ simple Monte Carlo simulations to explore the outcomes of two natal kicks on massive stellar binaries orbiting a MBH within $0.1$~pc. Inwards to this limit $\sim 1.5\times 10^5$, stars are expected to reside, estimated from the M-$\sigma$ relation \citep[e.g.,][]{Tremaine+02}. Thus, this limit is used to estimate the effects of natal kicks on the system close to the MBH.

\subsection{Monte Carlo Birth Distributions}\label{sec:MC}
We run a total of 300,000 Monte Carlo simulations of a binary star (the ``inner binary") orbiting a MBH (the ``outer binary") in a hierarchical triple configuration system. The inner binary comprises of main sequence stars with birth masses $m_1$ and $m_2$, and the MBH has a mass of $m_{\rm MBH} = 4 \times 10^6~\msun$. The inner (outer) orbit is characterized by the following orbital parameters: the semi-major axis $a_1$($a_2$), the eccentricity $e_1$($e_2$), the argument of periapsis $\omega_1$ ($\omega_2$), the longitude of the ascending node $\Omega_1$ ($\Omega_2$), and the true anomaly $f_1$($f_2$). The inner and outer orbits have a mutual inclination of $i_{\rm tot}$. 

For our Monte Carlo simulations we set $m_1$ as the more massive star and therefore is always the one that undergoes the first supernova in the system. The birth value of $m_1$ is chosen from a Kroupa mass distribution ranging from $8-100~\msun$ \citep{Kroupa01}, and $m_2/m_1= q$, where the mass ratio $q$ is chosen from a uniform distribution ranging from $0.1 -1$. Note that because of this mass distribution, some $m_2$'s will be white dwarf (WD) progenitors, and a small percentage will not undergo SN at all. Among the systems that are not eliminated in the pre-SN evolution (see Section \ref{sec:preSN}), \NScount \% of stars become NSs, \BHcount \%  become BHs, WD \WDcount \% become WDs, and \STcount \% never become a compact object (CO). 

The inner (outer) orbital eccentricities are chosen from a uniform (thermal) distribution ranging from $0-1$. We distribute the mutual inclination $i_{\rm tot}$ isotropically. Further the arguments of periapsis, true anomalies, and the inner binary longitude of the ascending node $\Omega_1$ are chosen from a uniform distribution between $0$ and $2 \pi$. Since we are working in the invariable plane reference plane, $\Omega_2 = \Omega_1 - \pi$.

 The outer semi-major axis $a_2$ is chosen from a number density distribution that goes as $n \propto r^{-2}$ with a minimum value of $1000$~au, which is roughly the semi-major axis of the well-studied star S0-2 in the S-cluster in the Galactic Center, and a maximum value of $0.1$~pc (see above). Note that current observations of late-type giant stars in the Galactic Center show a shallow core-like profile, rather than a cusp \citep[e.g.][]{Do+13}. It remains debated whether this flat density distribution is representative of the rest of the stellar population \citep[e.g.][]{Genzel+03,Do+09,Bartko+10,Schodel+20}. Theoretical models have found steeper cusps ranging from 1.5 - 2.75, depending on the mass component \citep[e.g.]{Bahcall+77,AH09}. We expect the density distribution of binaries around the MBH to have an effect on some parts of our results. For example, we expect more EMRIs for a steeper cusp since supernovae closer to the MBH are more likely to result in an EMRI \citep{Bortolas+Mapelli}. On the other hand, we do not expect that our results for X-ray binaries will be greatly affected, since their formation depends more on the properties of the inner binary. We leave a comprehensive exploration of the effect of varying the density distribution to a future work. 
 
 We choose the inner binary semi-major axis $a_1$ from the period distribution $dn/dP \propto {\rm log}(P)^{-0.45}$ \citep{Sana+13}, with the minimum and maximum values of $a_1$ chosen for each system according to the following criteria. First we require the inner orbit pericenter to be greater than 2 times the Roche limit of the system:
\begin{equation}\label{eqn:pericenter_Roche}
    a_1 (1 - e_1) > 2~a_{\rm Roche}.
\end{equation}
The Roche limit $a_{{\rm Roche},ij}$ for either star in the inner binary\footnote{For Equation \ref{eqn:pericenter_Roche}, we take $a_{\rm Roche} = {\rm max}[a_{{\rm Roche},12},a_{{\rm Roche},21}]$} is defined as:
\begin{equation}\label{eq:aRoche}
    a_{{\rm Roche},ij} = \frac{R_j}{\mu_{{\rm Roche},ji}},
\end{equation}
where $R_j$ is the radius of the star of mass $m_j$, found by the ZAMS mass radius relation:
\begin{equation}\label{eq:radius}
R(m) = 1.01~m^{0.57}~\rsun  
\end{equation}
\citep{Demircan+Kahraman}, and $\mu_{{\rm Roche},ji}$ is defined by:
\begin{equation}
\mu_{{\rm Roche},ji} = \frac{0.49(m_j/m_i)^{2/3}}{0.6 (m_j/m_i)^{2/3} + {\rm ln}(1 + (m_j/m_i)^{1/3})}
\end{equation}
\citep{Eggleton83}. To find the upper limit of $a_1$, we require that each triple system is dynamically stable \citep[][]{NaozReview}:
\begin{equation}
    \frac{a_1}{a_2}\frac{e_2}{1 - e^2_2} < 0.1 
\end{equation}

\begin{figure*}
    \centering
    \includegraphics[width=\linewidth]{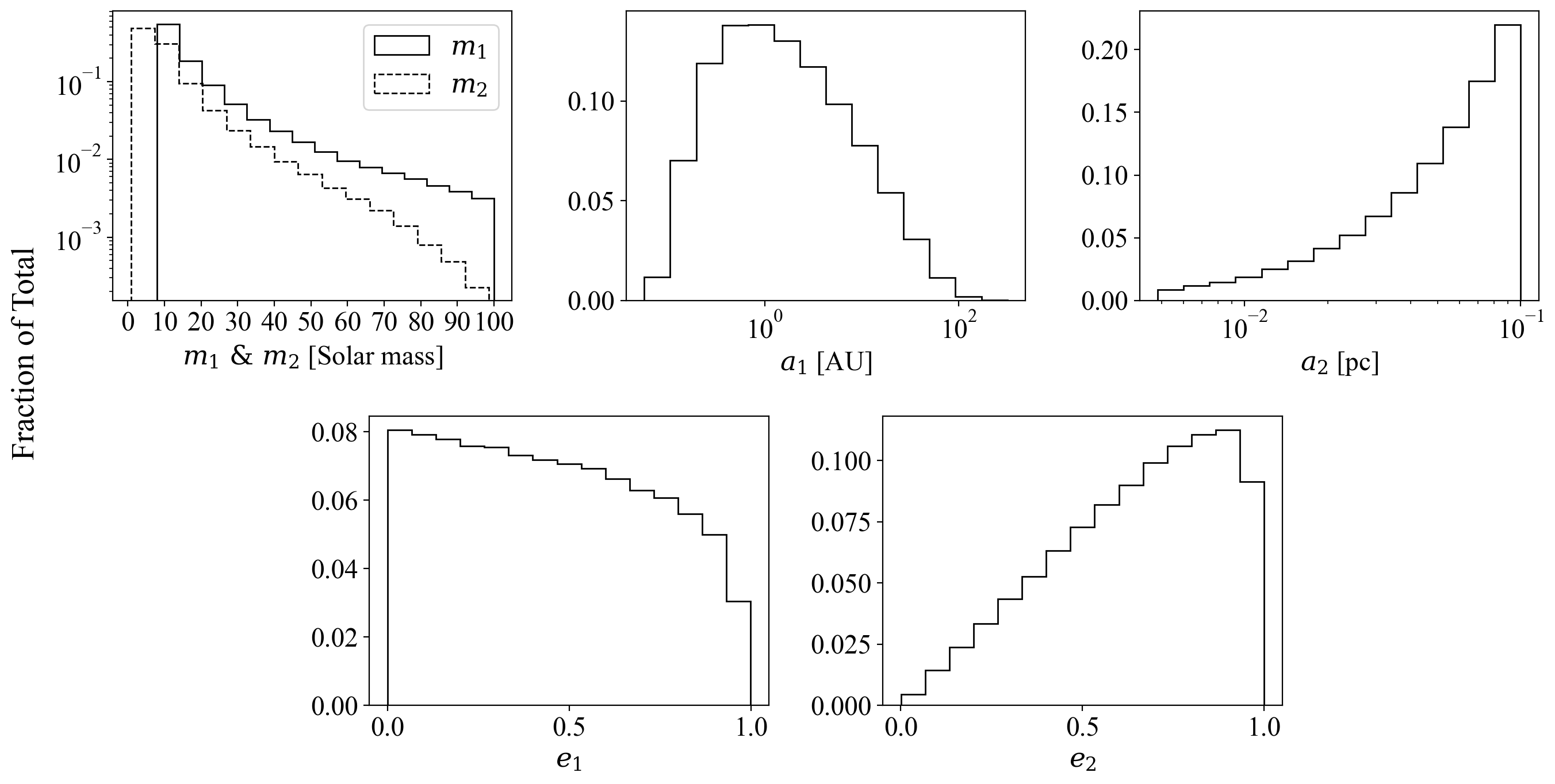}
    \caption{Distributions of the initial triple parameters.}
    \label{fig:ICs}
\end{figure*}

Finally, we require that the inner binary does not cross the Roche limit of the central MBH, and thus eliminate systems that do not fulfill the criterion below:
\begin{equation}\label{eq:RocheCrossing}
    a_2 (1 - e_2) > a_1 (1 + e_1)\Big(\frac{3~m_{\rm MBH}}{m_1 + m_2}\Big)^{1/3}.
\end{equation}
\citep[e.g.][]{Naoz+Silk}. We show the initial distributions of $m_1$, $m_2$, $a_1$, $a_2$, $e_1$, and $e_2$ in Figure \ref{fig:ICs}.

\subsection{Pre-SN Evolution}\label{sec:preSN}
For each star in the inner binary, we find the time at which it becomes a compact object and its mass before and after this transition with {\tt SSE} \citep{Hurley+00}. The stars experience mass loss due to main sequence evolution between birth and the first supernova, and between the first supernova and the second supernova. Between birth and the first supernova the inner binary will expand due to mass loss (the outer binary will also expand, but by a negligible amount due to the large mass of the MBH). We find the inner binary semi-major axis immediately before the first supernova, $a_{1,{\rm pre-SN}}$ by adopting adiabatic expansion, which conserved specific angular momentum.:
\begin{equation}\label{eq:massloss}
a_{1,{\rm pre-SN}} = \frac{m_1 + m_2}{m_{1,{\rm pre-SN}} + m_{2,{\rm pre-SN}}} a_1 \ .
\end{equation}

We note that we do not perform a numerical integration of the triple system either before the first kick or between the first and second kick. Eccentricity oscillations in the inner binary driven by the Eccentric Kozai-Lidov effect \citep[EKL,][]{Naoz16}, coupled with stellar evolution, can cause the inner stellar binary to become tidally locked, collide, or overflow its Roche Lobe, long before either of the stars undergo a supernova \citep[e.g.,][]{Stephan+16,Stephan+19}. We cannot account for all of these systems without a numerical simulation. However, we use an analytical criteria to identify systems where EKL causes the stellar binary to overflow its Roche Lobe. The EKL timescale on which the inner binary experiences eccentricity oscillations is approximately \citep[e.g.,][]{Antognini15}:
\begin{equation}\label{eq:tEKL}
   t_{\rm EKL} \sim \frac{16}{15}\frac{a^3_2 (1 - e^2_2)^{3/2} \sqrt{m_1 + m_2}}{a^{3/2}_1 m_{\rm MBH} k} \ .
\end{equation}
This timescale needs to be shorter than the inner binary General Relativity (GR) precession timescale $t_{\rm GR}$ in order for EKL to excite the inner binary eccentricity\footnote{However, in some cases, when the GR timescale is comparable or longer than the EKL timescale, the GR precession can trigger eccentricity excitation even in regimes that supress the large eccentricity excitations \citep[e.g.,][]{Naoz+2013,Hansen+20}.}. The GR precession timescale of the inner binary is defined as:
\begin{equation}\label{eq:tGR}
    t_{\rm GR, inner} \sim 2 \pi \frac{a^{5/2}_1 c^2 (1 - e^2_1)}{3 G^{3/2} (m_1 + m_2)^{3/2}} \ ,
\end{equation}
where $c$ is the speed of light. 

For systems with $t_{\rm ELK} < t_{\rm GR, inner}$ and mutual inclination between $40^{\circ}$ and $140^{\circ}$ the maximal eccentricity $e_{1, {\rm max}}$ achieved by the inner binary during one Kozai-Lidov cycle can be approximated analytically \citep[e.g.][]{Miller+Hamilton,Wen,Fabrycky+Tremaine,Liu+15}. We approximate $e_{1, {\rm max}}$ using the method outlined in \citet{Wen}.

If the minimum pericenter distance achieved during $t_{\rm EKL}$ (we also require that $t_{\rm EKL}$ is less than the timescale to the first SN) is less than the binary Roche Limit $a_1 (1 - e_{1,{\rm max}}) < a_{\rm Roche}$, the stellar binary will undergo Roche Lobe overflow and mass transfer. We found that roughly 10\% of initial stellar binaries fit these criteria. The subsequent binary evolution is out of the scope of our methodology. Thus, we eliminate these systems from our simulations and do not evolve them to the supernova stage. We note that the approach here uses the three-body Hamiltonian up to the quadrupole order, and cannot account for extreme eccentricities excited by the octupole term \citep[e.g.][]{NaozReview}. Moreover, for inner and outer eccentric orbits, the EKL mechanism can excite high eccentricity for wide range of inclinations \citep[e.g.,][]{LN,Naoz+11sec,Li+13,Li+14Chaos}. Therefore, our approach here eliminates conservative fraction of systems. However, we do expect that the binary orbital distribution from this approach will capture the distribution of the binaries post EKL, just before the first SN  \citep[as was highlighted in][]{Rose+19}.

We note that even though we eliminate these systems in the pre-SN stage, the statistics that we give later in the paper are with respect to the initial triple population we generated in Section \ref{sec:MC}. This is so that the event formation fractions we give later on in the paper can be simply multiplied by a star formation rate.

\subsection{Supernova Kicks}\label{sec:SNkicks}
We then apply the supernova kick to $m_1$.
We assume supernova kicks that are isotropically distributed in the inner binary orbital frame. We also assume instantaneous mass loss at the moment of the SN, using post-SN masses found with {\tt SSE} \citep{Hurley+00}. For NS progenitors, we assume kicks drawn from a normal distribution with an average of 400 km/s and standard deviation of 265 km/s \citep[e.g.,][]{Hansen+Phinney97,Arzoumanian+02,Hobbs+04}. For WDs, which are present in the second set of kicks, we assume a normal distribution with an average of 0.8 km/s and standard deviation of 0.5 km/s \citep[e.g.,][]{ElBadry+Rix,Hamers+Thompson}. There is much uncertainty about whether BHs receive natal kicks, and if they do, what the distribution of kick velocities would be \citep[e.g.,][]{Gualandris+05BHkick,Willems+05,Fragos+09,Repetto+12}. Thus, we adopt three different kick distributions for BH progenitors:
\begin{enumerate}
    \item ``Fast" BH kicks that are drawn from the same distribution as NS kicks.
    \item Following \citet{Bortolas+Mapelli}, ``slow" BH kicks that assume BHs receive the same linear momentum as NSs (i.e. we draw a kick velocity from the NS distribution, then multiply this by ($1.4~ \msun/m_{\rm BH}$), where $m_{\rm BH}$ is the mass of the BH progenitor, and 1.4 $\msun$ is the typical NS mass).
    \item No BH kicks.
\end{enumerate}

\begin{figure*}[!htb]
    \centering
    \includegraphics[width = \linewidth]{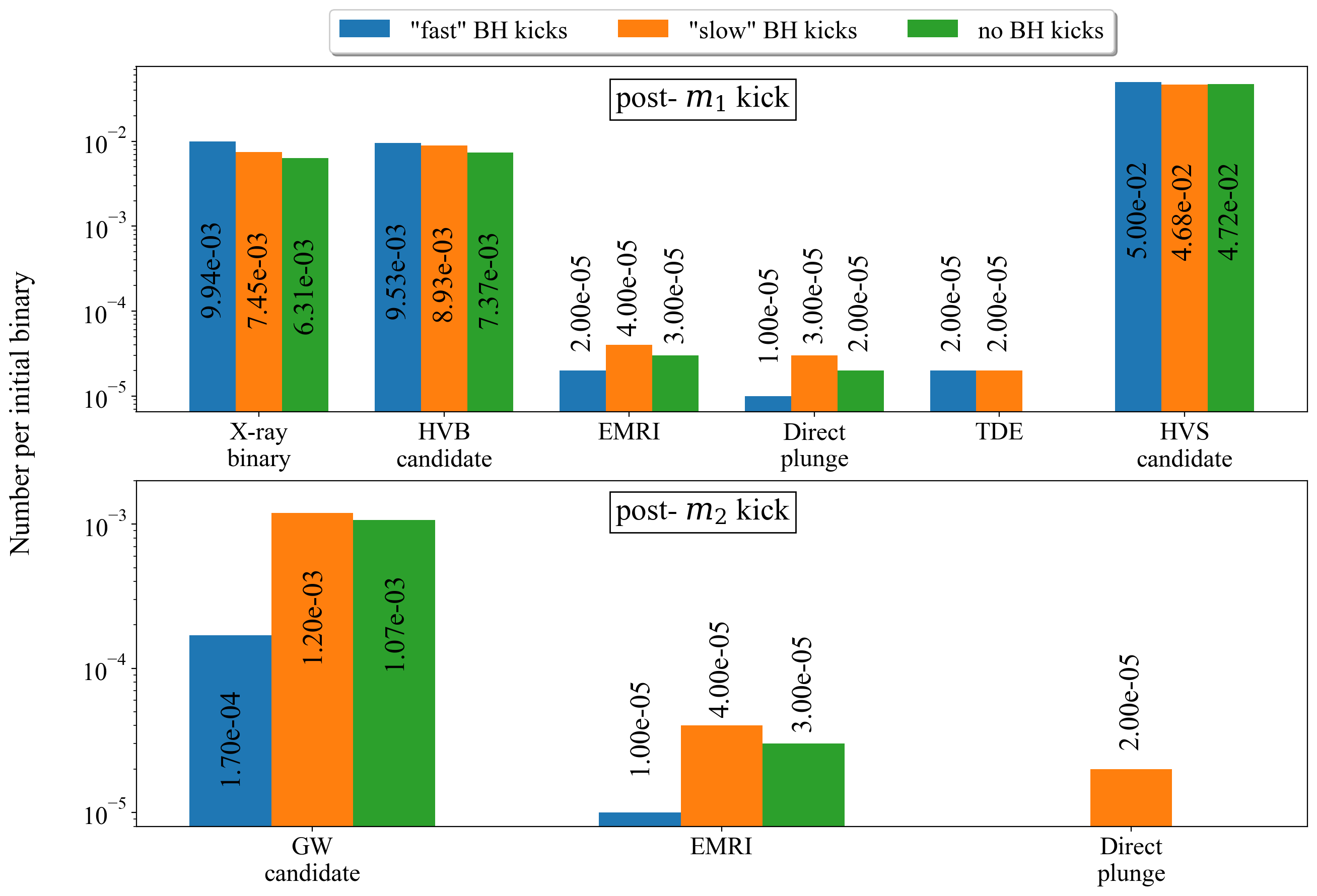}
    \caption{\textbf{Number of observables per initial stellar binary}\textsuperscript{\ref{ratecalc}}, triggered by the first natal kick (top panel), and by the second natal kick (bottom panel). For details about how we classify the different observables, see the following sections: X-ray binaries (\ref{sec:XrayBin}); HVB candidates (hypervelocity binaries, \ref{sec:HVB}); GW candidates (\ref{sec:GWmergers}); HVS candidates (hypervelocity stars, \ref{sec:HVS}); EMRIs (extreme mass ratio inspirals, \ref{sec:EMRI}); direct plunges (\ref{sec:Plunge}); and TDEs (tidal disruption events, \ref{sec:TDE}).}
    \label{fig:AggregateStats}
\end{figure*}

We run 100,000 simulations for each scenario, for a total of 300,000 simulations. We present statistics from all three scenarios in Sections \ref{sec:single_outcomes} and \ref{sec:binary_outcomes}, and in Figure \ref{fig:AggregateStats}.

 To apply the the natal kick, we simply add the Cartesian velocity kick vector to the velocity vector of $m_1$, and change $m_1$ to the post-SN mass found with SSE. We then use new velocity vector and mass to recalculate new orbital parameters. 
\begin{table}
\makegapedcells
\begin{tabular}{|P{2cm}||P{1cm}|P{1cm}|P{1cm}|P{1cm}|}
\hline
\multirow{2}{2cm}{\textbf{BH kick distribution}}
& \multicolumn{2}{P{2.365cm}|}{\textbf{Inner binary disrupted (\%)}}
& \multicolumn{2}{P{2.365cm}|}{\textbf{Inner binary survived (\%)}}     \\
\cline{2-5}
& post-$m_1$ kick
& post-$m_2$ kick
& post-$m_1$ kick
& post-$m_2$ kick
\\
\hline \hline 
fast 
& 88.2  & 94.5 & 11.8 & 5.5 \\
\hline
slow
& 73.0 & 88.1 & 27.0 & 11.9 \\
\hline
none
& 66.7 & 80.1 & 33.3 & 19.9 \\
\cline{1-5}
\end{tabular}
\caption{\label{tab:innerbinary} \textbf{Percentage of inner binaries disrupted/ survived} after $m_1$'s and $m_2$'s kick. As expected, the percentage of inner binaries that survive both kicks is inversely correlated to the magnitude of BH natal kicks.}
\end{table}

 The two main scenarios resulting from this first kick are the \textbf{``inner binary survived"} and the \textbf{``inner binary disrupted"} cases (see Figure \ref{fig:schematic}). In the case that the inner binary survives the first kick, we have two further branching scenarios. The inner binary can either stay bound to the MBH (ellitical orbit), or become unbound from the MBH (hyperbolic orbit). In the case that the inner binary is disrupted by the first kick, the component masses form separate binaries with the MBH (i.e. $m_1$-MBH and $m_2$-MBH). These binaries can either be bound or unbound. We then evolve $m_2$ up to its own natal kick, and recalculate the orbits after $m_2$'s kick. Between $m_1$'s and $m_2$'s kicks we evolve the orbits as follows:
 \begin{enumerate}
     \item We adiabatically expand the orbits, due to $m_2$'s mass loss, using Equation \ref{eq:massloss}
     \item For hyperbolic orbits, we solve the hyperbolic Kepler Equation using the HKE-SDG package \citep{HKE-SDG} to find the hyperbolic anomaly at the time of $m_2$'s kick, from which we calculate the true anomaly. Using the newly calculated true anomaly with the Keplerian elements of the orbit, we calculate the Cartesian coordinates of the objects in the orbit.
     \item For elliptical orbits, how we calculate the true anomaly at the time of $m_2$'s kick depends on the timescales of the system. If the timescale between the first and second natal kick is longer than 10 times the orbital period, we simply choose the eccentric anomaly at the time of $m_2$'s kick randomly from a uniform distribution between 0 and $2\pi$. Otherwise, we solve the elliptical Kepler's Equation iteratively using Newton's Method to find the eccentric anomaly. From this we calculate the true anomaly, and then the Cartesian coordinates of the orbit. 
 \end{enumerate}
 
 After each kick we check whether the natal kick had induced any observable phenomena (e.g., X-ray binary, GW merger, etc.). We describe our criteria in Section \ref{sec:single_outcomes} for the cases where the inner binary is disrupted, and in Section \ref{sec:binary_outcomes} for the cases where the inner binary survived. We also present results and statistics from our simulations for each outcome in the the following two sections. We include a simplified schematic of our process in Figure \ref{fig:schematic}.

\begin{figure*}
    \centering
    \includegraphics[width=\linewidth]{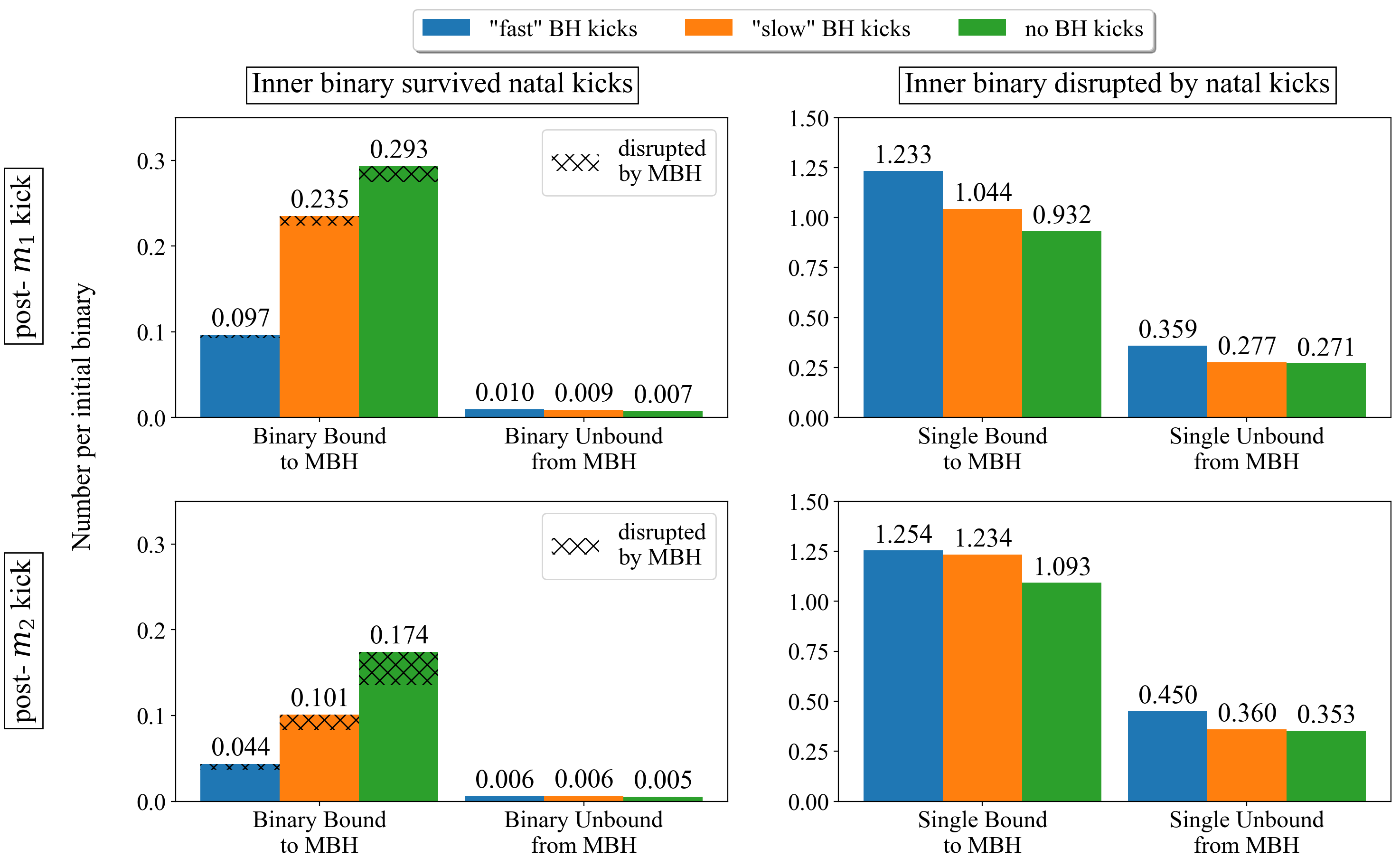}
    \caption{\textbf{Summary of orbital configuration resulting from each natal kick}. In the top (bottom) panels we show the results from after the first (second) natal kick. In the left (right) panels we show the orbital configurations where the inner binary survived (is disrupted by) the natal kicks. In the left panels, we use the crossed pattern to represent the binaries that survived the natal kick, but will be disrupted by crossing the MBH Roche limit upon pericenter passage (Section \ref{sec:bin_disruption}). We note that for the ``Single Bound to MBH" cases, we have numbers greater than 1 because this number is taken relative to number of initial binaries, and each binary contains two singles.}
    
    \label{fig:orb_config}
\end{figure*}

\section{Inner Binary Disrupted by Natal Kicks}\label{sec:single_outcomes}

In the case that the inner binary is disrupted by either $m_1$'s or $m_2$'s natal kick, the result is two separate binaries with each composed of a stellar mass object orbiting the MBH ($m_1-{\rm MBH}$ and $m_2-{\rm MBH}$). These orbits can either be bound (elliptical orbit), or unbound (hyperbolic orbit). However, natal kicks are less likely to result in unbound objects, as we find that the majority of objects remain bound to the MBH. See Table \ref{tab:singles}, for percentages of bound and unbound orbits as a function of stellar type. 

The orbital configurations and observables resulting from these systems is described below. We also give rates of each orbital configuration and observables from our Monte Carlo simulations. These rates are summarized in bar chart form in Figures \ref{fig:orb_config} and \ref{fig:AggregateStats}, which also break down the rates by BH kick distribution.

\begin{table*}
\centering
\makegapedcells
\begin{tabular}{|P{2cm}||P{1.5cm}|P{1.5cm}|P{1.5cm}|P{1.5cm}|P{1.5cm}|P{1.5cm}|}
\hline
\multirow{2}{2cm}{\textbf{BH kick distribution}}
& \multicolumn{2}{P{3.365cm}|}{\textbf{NS (\%)}}
& \multicolumn{2}{P{3.365cm}|}{\textbf{BH (\%)}} 
& \multicolumn{2}{P{3.365cm}|}{\textbf{WD (\%)}}     \\
\cline{2-7}
& bound
& unbound
& bound
& unbound
& bound
& unbound
\\
\hline \hline 
fast 
& 67.1 & 32.9 & 67.7 & 32.3 & 91.7 & 8.3\\
\hline
slow
& 65.9 & 34.1 & 96.3 & 3.7 & 91.5 & 8.5 \\
\hline
none
& 65.7 & 34.3 & 98.3 & 1.7 & 91.1 & 8.9\\
\cline{1-7}
\end{tabular}
\caption{\label{tab:singles} \textbf{Percentage of single compact object orbits that are bound to/unbound from MBH} after both kicks, for different types of compact objects. Whereas majority of single BHs and WDs remain bound to the MBH after the natal kicks (except for the case where BHs receive ``fast" kicks), about a third of single NSs are unbound from the MBH after natal kicks.}
\end{table*}

\subsection{Single objects bound to MBH}\label{sec:single_bound}
\begin{figure*}[!htb]
    \centering
    \includegraphics[width = \linewidth]{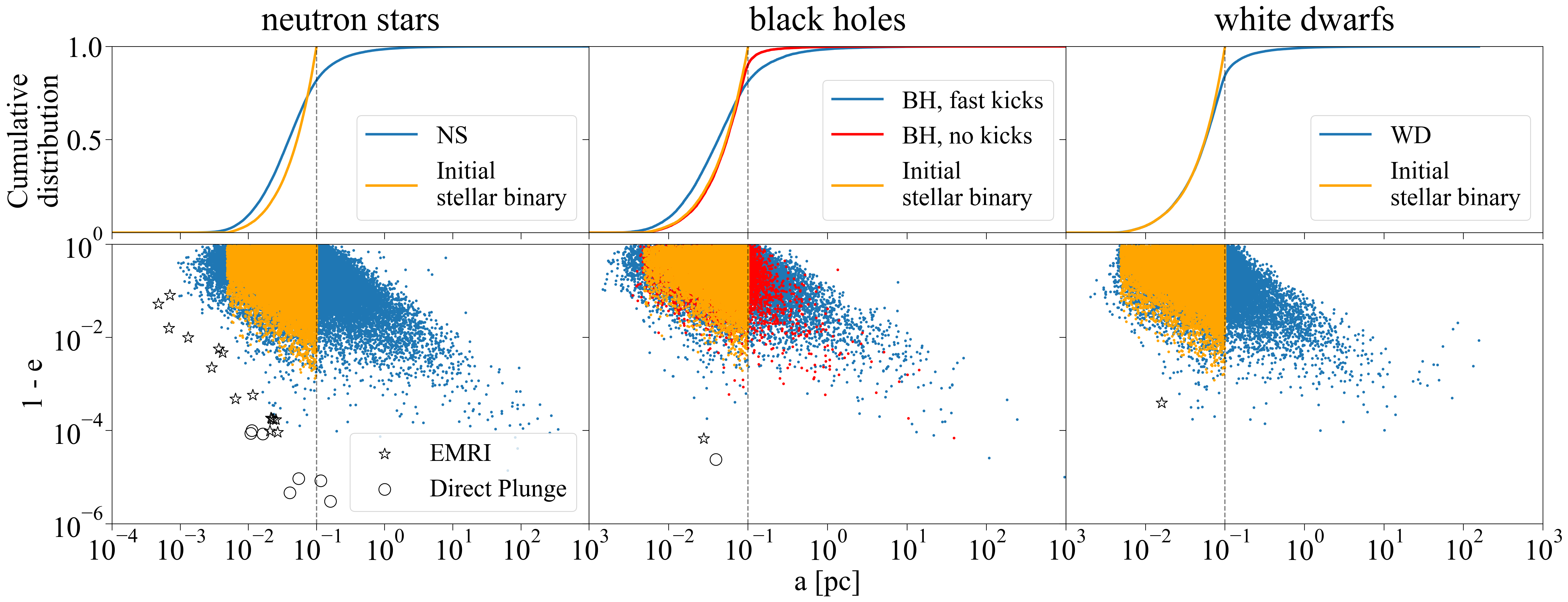}
    \caption{\textbf{Distribution of bound single compact object orbits after two successive natal kicks} in the  $(a, 1 -e)$ plane ($a$ and $e$ are the semi-major axis and eccentricity with respect to the MBH). These objects were components of the stellar binaries that were disrupted by the natal kicks. For the NSs and WDs, we show data only from the ``fast" BH kicks simulation set as the NS and WD kick distributions do not change between simulation sets and the results from any one set is fairly representative of the others. For the BHs, we show data from both the ``fast" (in blue) and no BH kicks (in red) simulation sets to demonstrate two extremes. We plot the distribution (with respect to the MBH) of the initial stellar binaries that contained the compact object progenitors in yellow for comparison. In addition, we plot the distribution of EMRI progenitors (empty star symbol) and direct plunge progenitors (empty circles).}
    \label{fig:BoundSingles}
\end{figure*}

Single objects bound to the MBH is the most common resulting orbital configuration in our simulations. We find 0.9-1.2 single objects per initial binary bound to the MBH after the first natal kick, and 1.1-1.3 after both natal kicks (see Figure \ref{fig:orb_config}). The vast majority of these orbit are long term stable orbits, with a very small number becoming EMRIs (Section \ref{sec:EMRI}), direct plunges (Section \ref{sec:Plunge}), and TDEs (Section \ref{sec:TDE}).

In Figure \ref{fig:BoundSingles} we show the distribution of NSs, BHs, and WDs with respect to the MBH after two successive natal kicks. Note that each triple is evolved in its own simulation, and each system undergoes natal kicks at different times, thus Figure \ref{fig:BoundSingles} does not accurately show the distribution of a contemporaneous population of COs. However, Figure \ref{fig:BoundSingles} is a good approximation of NS and BH distributions between a few 10 Myrs and $\sim 1 {\rm Gyr}$ after the star formation episode. This is because NS natal kicks occur between $\sim 10-45$ Myr and BH natal kicks occur between $\sim 4-10$ Myr, and the two-body relaxation timescale in the Galactic Center is $\sim 0.1 -1~{\rm Gyr}$ \citep[][]{Rose+20}. WD natal kicks occur between 45 and $1.4 \times 10^4$ Myr, so Figure \ref{fig:BoundSingles} cannot be taken as representation of a comtemporaneous population of WDs at any given time.

Figure \ref{fig:BoundSingles} shows that after two successive natal kicks, the cusp of single NSs is steeper than the initial stellar binary cusp. This is also true for the BHs, if we assume fast BH kicks. WDs and BHs with no kicks more or less follow the initial stellar binary distribution. 

\begin{figure*}
    \centering
    \includegraphics[width=0.9\linewidth]{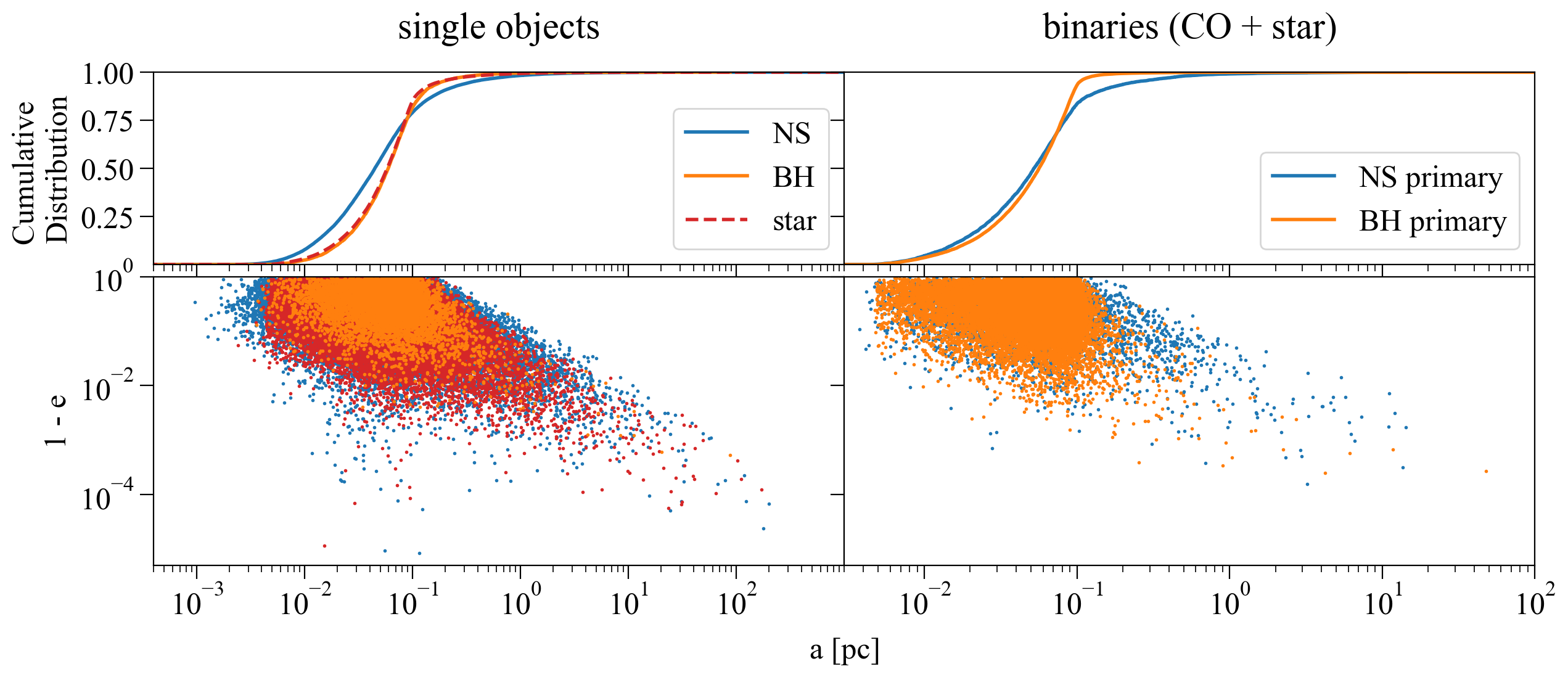}
    \caption{\textbf{Distribution of single objects and binaries bound to the MBH after the first natal kick} in the  $(a, 1 -e)$ plane (bottom panels), and the cumulative distribution of $a$ (top panels), for the slow BH kicks scenario. The variables $a$ and $e$ are the semi-major axis and eccentricity with respect to the MBH.}
    \label{fig:pk1_bound}
\end{figure*}

In fact, a steeper NS cusp is already present after just one natal kick. In Figure \ref{fig:pk1_bound} we show the distribution of single NSs, BHs, stars, and binaries bound to the MBH after the first natal kick, for the slow BH kicks scenario.  We see a steeper disribution of NSs than for BHs and stars, and a steeper distribution of binaries with a NS primary than for binaries with a BH primary. Thus, for a relatively young stellar population, before mass segregation has taken place, we expect a steeper cusp of NSs relative to the stellar and BH cusp. This is of course excepting the case in which BHs receive fast kicks similar to NSs, in which case the BH cusp will be as steep as the NS cusp after the  first natal kick (not shown in figure).

\subsubsection{Extreme Mass Ratio Inspirals (EMRIs)}\label{sec:EMRI}
Extreme Mass Ratio Inspirals (EMRIs) are gradual GW inspirals of stellar-mass COs onto MBHs, and are one of the main targets for the future Laser Interferometer Space Antenna (LISA) \citep{LISAL3}. The timescale on which a CO inspirals onto an MBH is \citep{Peters64}:
\begin{equation}
    t_{\rm GW, EMRI} \sim \frac{5}{64}\frac{c^5 a^4}{G^3 m^2_{\rm MBH} m}(1 - e^2)^{7/2} \ .
\end{equation}
In a dense environment such as the Galactic Center, we also have to take into account the effects of two-body relaxation on the potential EMRI. While two-body relaxation from other stars in the cluster can create EMRIs by pushing COs onto high eccentricity orbits, it can also push COs onto more circular orbits, such that $t_{\rm GW, EMRI}$ increases to longer than a Hubble time. Two-body relaxation changes an orbit's angular momentum by an order of itself on a timescale of $\sim (1 - e)t_{\rm relax}$, where $t_{\rm relax}$ is:
\begin{equation}
 t_{\rm relax} = \frac{0.34}{{\rm ln} \Lambda}\frac{\sigma^3(r)}{G^2 \rho(r) m},
\end{equation}
where $\rm ln~\Lambda = 15$ is the Coulomb algorithm, $\sigma(r)$ is the velocity dispersion, given by \citep{Kocsis+11}:
\begin{equation}\label{eq:sigma}
    \sigma(r) = 280~{\rm km~s^{-1}}~\sqrt{\frac{0.1~{\rm pc}}{r}},
\end{equation}
and $\rho(r)$ is the Galactic Center stellar density \citep{Genzel+10}:
 \begin{equation}\label{eq:rho}
     \rho(r) = 1.35 \times 10^6~\msun~{\rm pc}^{-3} (\frac{r}{0.25~{\rm pc}})^{-1/3} \ .
 \end{equation}
Thus, we require that $t_{\rm GW,EMRI} < (1 - e) t_{\rm relax}$ to be classified as an EMRI \citep[e.g.,][]{EMRIReview}.

EMRI formation is a rare event in our simulations. In particular, we find \EMRI\ EMRIs per initial stellar binary\footnote{\label{ratecalc}Here we remind the reader that we generated 100,000 initial conditions per set of Monte Carlo simulations. To arrive at the ``EMRIs (or any other outcome) per initial stellar binary" figure, we simply count the number of EMRIs (or any other outcome) in each set of simulations and divide this by 100,000.}. This is roughly consistent with results from \citet{Bortolas+Mapelli}, which studied EMRIs triggered by natal kicks on single stars in the Galactic Center and and found a rate of $10^{-7} - 10^{-4}$ EMRIs per SN. Assuming a star formation rate of $10^3~\msun {\rm Myr}^{-1}$ in the Galactic Center \citep[e.g.][]{Lu+09}, and an average binary mass of $30 \msun$ (from our Monte Carlo simulations), we obtain an EMRI rate of $(1 - 2.7) \times 10^{-9}~{\rm yr}^{-1}$. Assuming a galaxy density of $0.02~{\rm Mpc}^{-3}$ \citep{Conselice+2005}, and that half of galaxies contain a MBH at the center, we calculate an EMRI volume rate of $0.01-0.027~{\rm Gpc}^{-3}~{\rm yr}^{-1}$ from our simulations\footnote{Note that these rates are about one to two orders of magnitude lower than the EMRI rate estimated via two-body relaxation processes \citep[e.g.,][]{Hopman+Alexander06}, and three to four orders or magnitude lower than the estimated rate of MBH binaries \citep{Naoz+22}. }. However, we note that this rate is likely a conservative one because of the minimum value of 1000 AU that we have chosen for the binary distribution around the MBH.  As shown by \citet{Bortolas+19}, stars closer to the MBH are much more likely than stars further away to end up as an EMRI after undergoing an SN. Thus, EMRI rates are very sensitive to the inner edge of the initial binary distribution. Currently, there several observed stars in the S-cluster with semi-major axes smaller than 1000 AU: S0–102/S55 with a semi-major axis of $\sim 900$ AU \citep{Meyer+12}, S62 with a semi-major axis of $\sim 700$ AU \citep{Peisker+20}, and S4711, with a semi-major axis of $\sim 600$ AU \citep{Peisker+20b}. \citet{Bortolas+19} found that a population of single stars located between 0.001 pc ($\sim 200$ AU) and 0.004 pc, modelled after the S-cluster stars, can generate up to a few $10^{-4}$ EMRIs per SN. Thus, if S0–102/S55, S62, and S4711 are indicative of a larger population of stars and stellar binaries inside of 1000 AU, then the EMRI rate we have found here may underestimate the true rate from this mechanism by about an order of magnitude.

Interestingly, the three different BH kick cases give very similar EMRI formation rates. This is because in our simulations most EMRIs involve a NS inspiraling into the MBH (see Figure \ref{fig:BoundSingles}). Out of EMRIs from all three BH kick cases, \NSemri\ \% are NS-EMRIs, \BHemri\ \% are BH-EMRIs, and \WDemri\ \% are WD-EMRIs.

Note that the WDs will not behave like a NS or BH EMRI because while we have a GW signal detectable by LISA while the WD is inspiralling, this signal will stop when the WD is tidally disrupted by the MBH and and replaced with an EM counterpart \citep[e.g.][]{Sesana+08}.

\subsubsection{Direct Plunges}\label{sec:Plunge}
Natal kicks can results in orbits with such a high eccentricity that they cause the orbiting object to plunge directly into the MBH instead of gradually inspiralling like EMRIs. Plunging orbits satisfy:
\begin{equation}
    e > 1 - 4\frac{R_s}{a},
\end{equation}
where $a$ and $e$ are respectively the semi-major axis and eccentricity of the post-kick orbit, and $R_s = 2 G m_{\rm MBH}/c^2$ is the Schwarzchild radius of the MBH \citep[e.g.][]{Amaro+07}. Unlike EMRIs, which emit detectable GWs for thousands of orbits, creating a coherent signal, plunges emit a brief 
burst of GW radiation \citep[e.g.,][]{Yunes+08,Berry+13}. Therefore direct plunges are not expected to be detected by LISA unless originating from the Milky Way Galactic Center \citep{Hopman+Alexander05}. We find, in our simulations, \Plunge\ direct plunges per initial stellar binary. The distribution of direct plunge-progenitors is shown in Figure \ref{fig:BoundSingles}.

\subsubsection{Tidal Disruption Events (TDEs)}\label{sec:TDE}
Tidal disruption events (TDEs) will happen when the pericenter of the $m_2-{\rm MBH}$ orbit drops below the MBH tidal radius as a result from natal kick of $m_1$. We also require that $m_2$ pass within the tidal radius before it undergoes its own natal kick to qualify as a TDE. 

The tidal radius is defined as:
\begin{equation}
r_t \sim R_* \Big(\frac{m_{\rm MBH}}{m_*}\Big)^{1/3},
\end{equation}
where $R_*$ is the radius of the star, given by Equation (\ref{eq:radius}), and $m_*$ is its mass \citep[e.g.][]{Rees88}. We also require that $r_t$ lies outside of the MBH Schwarzchild radius so that the star is not directly swallowed instead of being tidally disrupted. TDEs are a a rare outcome in our simulations. We find \TDE\ TDEs per initial stellar binary for the ``fast" and ``slow" kicks scenarios. 

\subsection{Single objects unbound from MBH}\label{sec:unbound_single}

\begin{figure*}
    \centering
    \includegraphics[width=\linewidth]{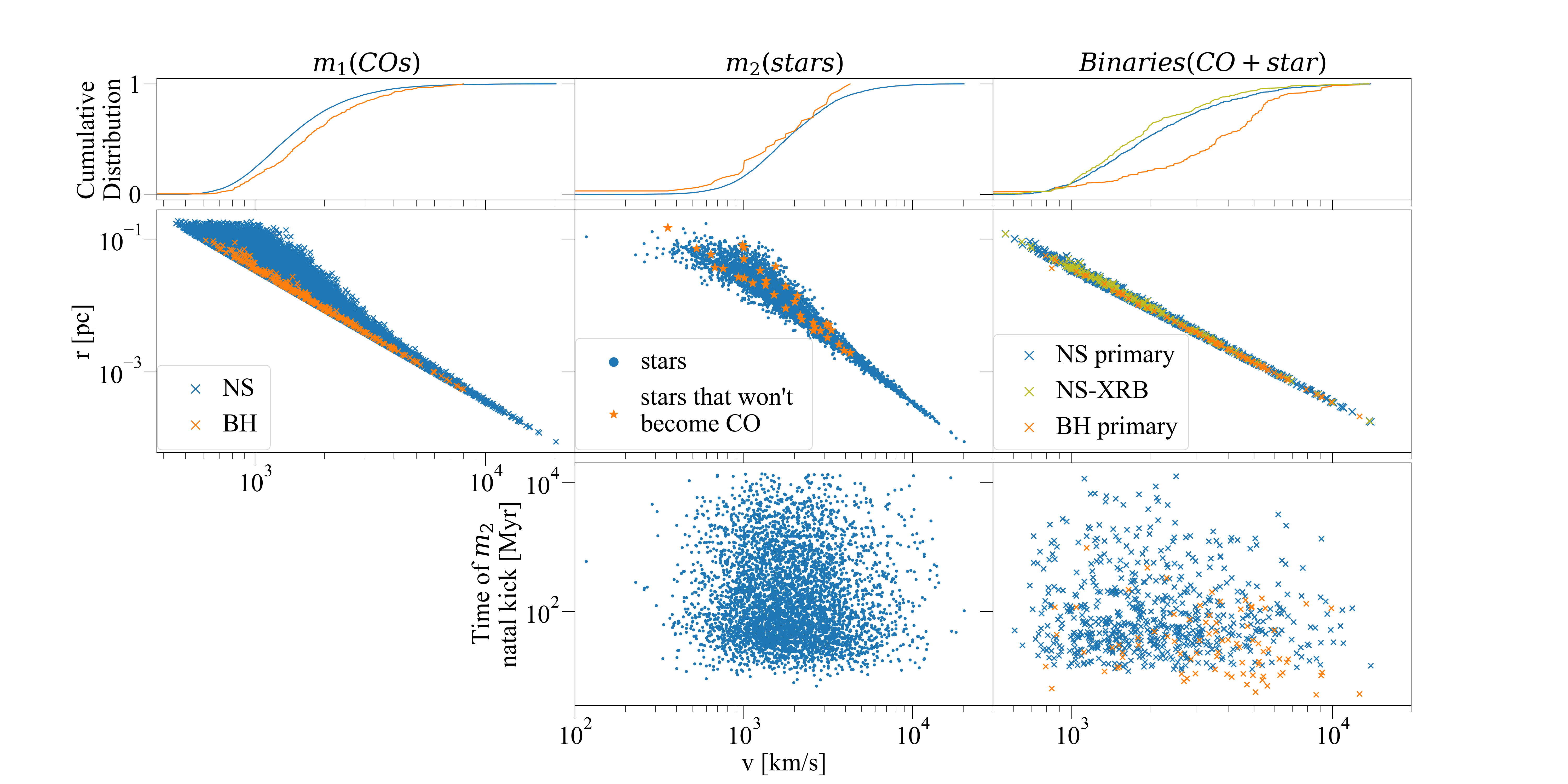}
    \caption{\textbf{Objects on unbound orbits after $m_1$'s natal kick} for the ``slow" BH kicks case. Here we show the \textit{immediately post kick} position and velocity of unbound objects (middle row), and the cumulative distribution of the velocities (top row). For the unbound stars and binaries, we also show the time at which $m_2$ will become a CO (bottom row). Note that this time is measured with respect to the initial stellar formation and not with respect to $m_1$'s SN. We also show the population of unbound NS X-ray Binaries in the third column.}
    \label{fig:HVS}
\end{figure*}

Natal kicks can result in objects that are unbound from the MBH (i.e.) on hyperbolic orbits. In our simulations we find 0.27-0.36 single objects per initial stellar binary are unbound from the MBH after the first kick, and 0.35-0.45 after both kicks, making this the second most common orbital configuration resulting from our simulations (see Figure \ref{fig:orb_config}). Most objects are ejected from the inner 0.1 pc with velocities between 1000-10,000 km/s in our simulations, as shown by Figure \ref{fig:HVS}. Despite deceleration by the combined gravitational potential of the MBH, nuclear cluster, and galaxy, objects ejected from the Galactic Center with velocities in excess of 1000 km/s should reach $r > 8$ kpc with ease, and retain their high velocities \citep[e.g.][]{Kenyon+08}. 

Unfortunately, hypervelocity NSs and BHs are not observable. However, a large number of $m_2$'s are ejected by $m_1$'s natal kick (see Figure \ref{fig:HVS}). Since $m_2 < m_1$ and thus all the $m_2$'s are still stars at the time of $m_1$'s kick, some $m_2$'s may be observed as hypervelocity stars. We discuss this possibility further in Section \ref{sec:HVS}.

\subsubsection{Hypervelocity stars}\label{sec:HVS}
If either the two members of the binary or one member of the binary becomes unbound to the MBH, it may be observed as a hyper-velocity star. 
A large number of stars have been detected with velocities larger than the the escape velocity from the Galaxy and with trajectories that are consistent with a Galactic Center origin \citep[e.g.,][]{Brown+05,Brown+07,Brown+09,Brown+09MMT,Brown+10,Brown+12,Brown+18,Hirsch+05,Kollmeier+10,Boubert+18}. 

The leading channel to explain these systems has involved a binary star crossing the tidal limit and breaks up, ejecting in the process one of the members \citep[known as the Hills mechanism,][]{Hills88}. The consequences of this mechanism have been explored in the literature in great details \citep[e.g.,][]{Gualandris+05,Bromley+06,Perets+07,Kenyon+08,Antonini+10,Sari+10,Kobayashi+12,Zhang+13,Rossi+14}. Additionally, it has been suggested that SN kicks could also contribute to hypervelocity population \citep[e.g.,][]{Zubovas+13,Lu+19,Bortolas+19}. In particular, \citet{Zubovas+13} predicted a characteristic spatial anisotropy for hypervelocity stars created by SN kicks. As highlighted in Figure \ref{fig:HVS}, we expect to have a population of hypervelocity stars as well as hypervelocity COs after the first natal kick. The first natal kick takes place between 4 and 45 Myr after the star formation episode, and as shown in the bottom row of Figure \ref{fig:HVS}, over half of the ejected stars will not become COs until 100 Myr to 10 Gyr after star formation. Combined with the fact that the majority of these stars are ejected from the inner 0.1 pc with velocities between 1000 and 10,000 km/s, a large fraction of them can be observed as hypervelocity stars outside of the Galactic Center.

\begin{figure*}[t]
    \centering
    \includegraphics[width=\linewidth]{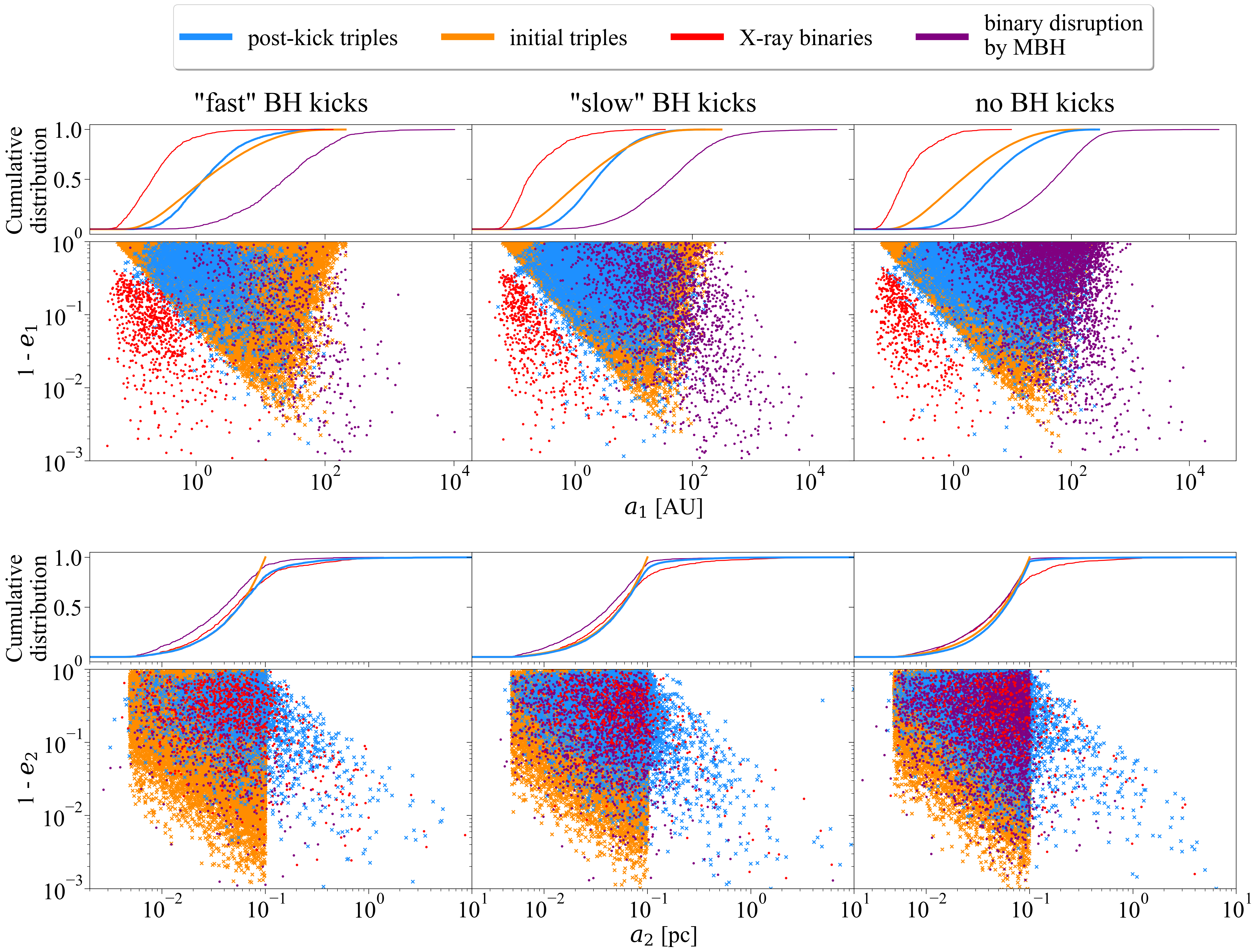}
    \caption{\textbf{Distribution of triples that remain after natal kicks} in the ($a_1$,$1-e_1$) plane (top plots, $a_1$ and $e_1$ are the inner binary semi-major axis and eccentricity), and the ($a_2$,$1-e_2$) plane (bottom plots, $a_2$ and $e_2$ are the outer binary semi-major axis and eccentricity). We show X-ray binaries created by $m_1$'s natal kick in red, stable triples in blue, and binary disruptions (these are binaries that survive the natal kicks but will be disrupted by the MBH at the next pericenter passage) in purple. A fraction of the binaries in the stable triples will become GW sources.}
    \label{fig:bound_binary}
\end{figure*}

\section{Inner Binary Survived Natal Kicks}\label{sec:binary_outcomes}
In the case that the inner binary survives $m_1$'s or $m_2$'s natal kick (or both), we have a stellar-mass binary orbiting the MBH. The orbital configurations and observables resulting from these systems is described below. We also give rates of each orbital configuration and observables from our Monte Carlo simulations. These rates are summarized in bar chart form in Figures \ref{fig:orb_config} and \ref{fig:AggregateStats}, which also break down the rates by BH kick distribution.

\subsection{Binaries bound to MBH}\label{sec:bin_bound}
We find 0.1-0.29 binaries per initial binary remain bound to the MBH after one natal kick, and 0.04-0.17 after both natal kicks. As highlighted in Figure \ref{fig:orb_config}, assuming no kicks for the BHs increases the survivable rate of binaries around the MBH. Furthermore, the number of BH-BH binaries after two kicks is inversely correlated with the strength of BH kicks. For the ``fast" BH kicks case, BH-BH comprise only 4.2\% of the survived binaries. This percentage increases to 39.5\% and 58.1\% for the ``slow" and no BH kicks case, respectively. As expected, due to high natal kick magnitudes, NS-NS binaries are relatively rare in our simulations. They comprise $\sim 1$\% of all surviving binaries for the fast BH kicks case, and $\sim 0.2$\% for the two other BH kick cases.
 
\begin{figure*}
    \centering
    \includegraphics[width=\linewidth]{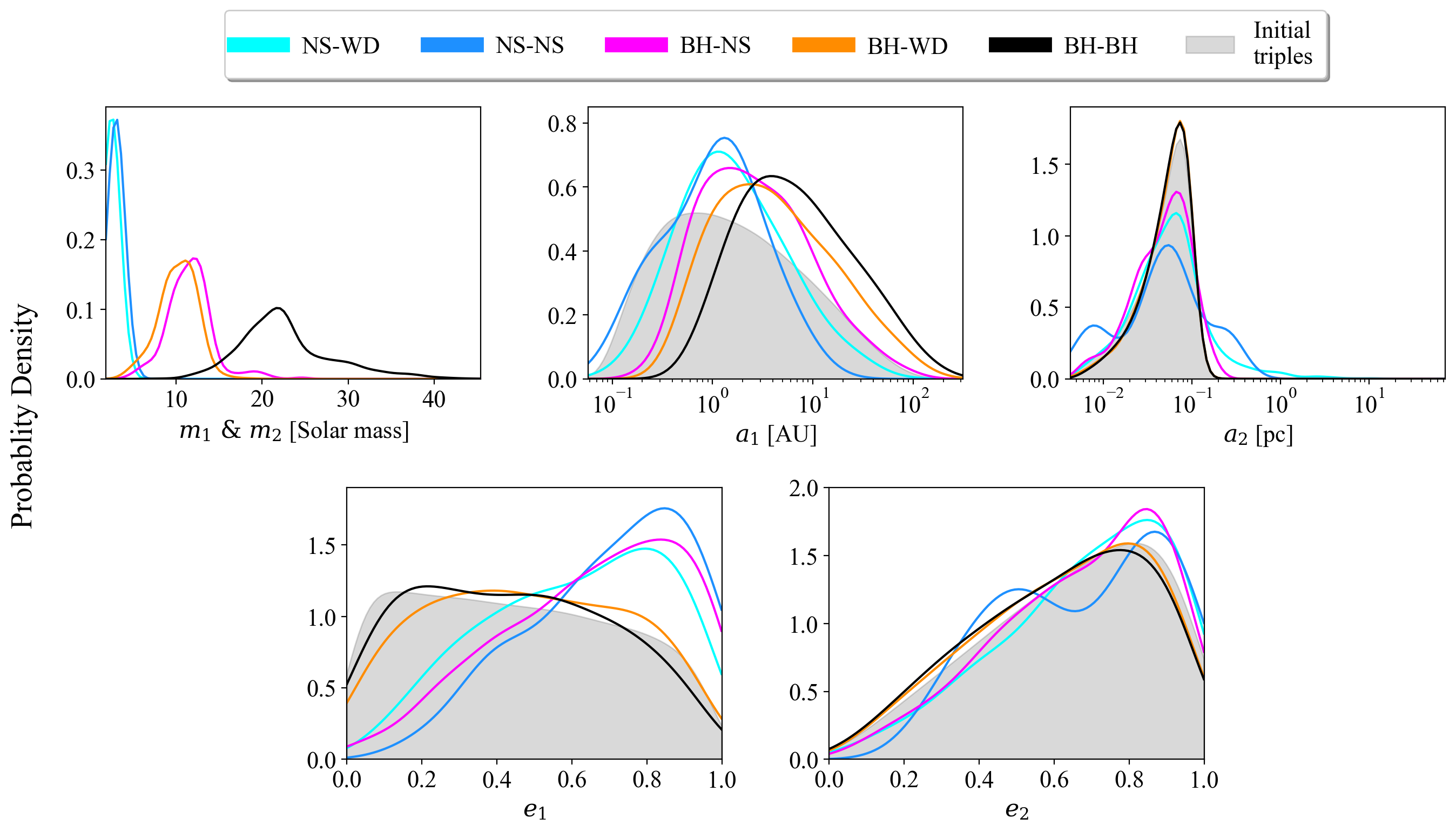}
    \caption{\textbf{Distributions of CO binaries bound to MBH after both natal kicks} for the no BH kicks case. The initial triple probability densities are shaded in gray for comparison. Notably, CO binaries containing at least one NS tend to be more eccentric than binaries with a BH primary due to the large NS kicks.}
    \label{fig:CO_binaries}
\end{figure*}

In Figure \ref{fig:CO_binaries} we show the distribution of orbital parameters for the surviving CO binaries for the no BH kicks case, compared to the initial binaries. We choose to highlight the no BH kicks because this is the case with the most surviving binaries. In general we find that the surviving binaries have larger semi-major axes ($a_1$) than the initial binaries. Surprisingly, despite the large NS kick magnitudes, the $a_1$ distribution for binaries containing at least one NS (i.e NS-WD, NS-NS, and BH-NS), peaks at a smaller value than for binaries that don't (i.e. BH-WD, BH-BH). This may be because binaries that start with a smaller $a_1$ have a better chance at surviving high NS natal kicks. Furthermore, the binaries containing at least one NS generally have much large eccentricities ($e_1$) than binaries that don't. We find similar trends with the semi-major axis and eccentricity of the binaries' orbits around the MBH ($a_2$ and $e_2$). In general, NS containing binaries have slightly larger $a_2$ and $e_2$ than the other types of CO binaries.

These surviving binaries can undergo EKL eccentricity excitations from the MBH which may result in GW mergers \citep[e.g.,][]{Hoang+18,Stephan+19}. We explore this effect in Section \ref{sec:GWmergers}. We also find X-ray binaries among the surviving binaries, which we discuss in Section \ref{sec:XrayBin}. 

\subsubsection{X-ray Binary}\label{sec:XrayBin}
Natal kicks can actually tighten the inner binary, causing its pericenter to drop below the binary Roche Limit, creating a X-ray binary. Note that we will only get X-ray binaries after $m_1$'s kick and not after $m_2$'s kick, since one member of the inner binary needs to be still a star. We classify a system as a potential X-ray binary if the inner binary post-kick pericenter is below $a_{\rm Roche}$, given by Equation (\ref{eq:aRoche}).
In our simulations we find \XrayBin\ X-ray Binaries per initial stellar binary, for the no BH kicks and ``fast'' BH kicks, respectively. 

\begin{figure*}[!htb]
    \centering
    \includegraphics[width=\linewidth]{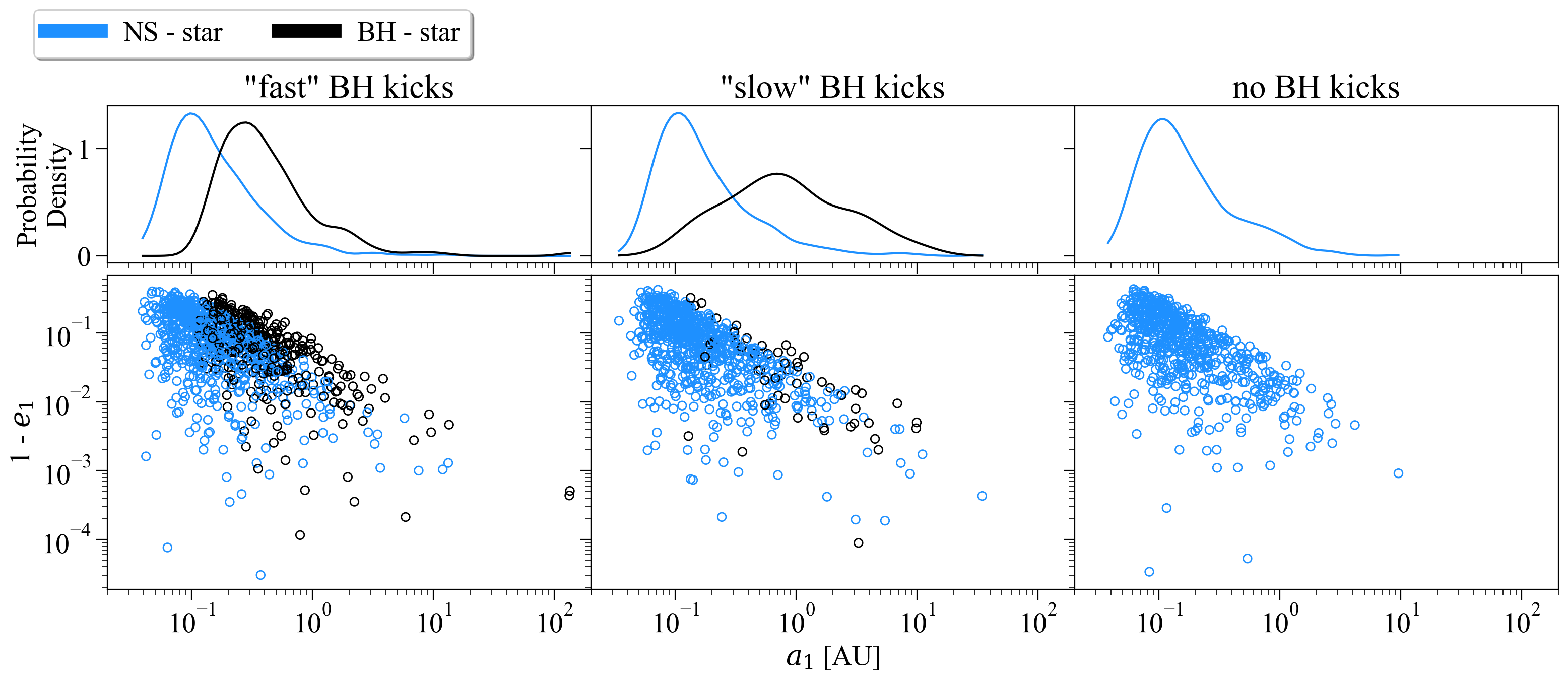}
    \caption{\textbf{Distribution of X-ray binaries} in the ($a_1$, $1 - e_1$) plane (bottom three panels), and the kernel density estimation (KDE) of $a_1$ (top three panels). X-ray binaries with a NS (BH) primary is shown in blue (black). We show X-ray binaries from all three BH kick distribution cases. We find that the set receiving ``fast" BH kicks resulted in the most BH-star X-ray binaries, with the set receiving no BH kicks resulting in no BH-star X-ray binaries at all. Furthermore, ``Fast" and ``slow" BH kicks result in different semi-major axis distributions for BH-star X-ray binaries, as shown by the KDEs in the top panels. Thus, the type and semi-major axis distribution of a statistically significant population of X-ray binaries in the GC can be a diagnostic for the hitherto uncertain distribution of natal kicks received by BH progenitors.}
    \label{fig:XrayBinary}
\end{figure*}

In Figure \ref{fig:XrayBinary} we show the distribution of X-ray binaries for the different kick models. Unlike Figure \ref{fig:bound_binary} here we divide the notation to BH X-ray binaries and NS X-ray binaries. As highlighted in Figure \ref{fig:XrayBinary}, the no BH kicks case has zero BH X-ray binaries. Thus, at face value, this implies that detecting BH X-ray binaries at the center of the galaxy may imply that BH may have some kicks\footnote{Note that other formation channels for X-ray binaries, such as EKL, \citep[e.g.,][]{Naoz+16,Stephan+19}, may form BH X-ray binaries even in the absence of BH kicks.  }.

Interestingly, our simulations predict some X-ray binaries binaries ejected from the GC (see Figure \ref{fig:HVS}). These comprise about 20\% of all X-ray binaries, for all BH kick scenarios. See Section \ref{sec:HVB} for further discussion.

\subsubsection{GW Candidates}\label{sec:GWmergers}
Natal kicks can shrink the inner binary orbit or increase the eccentricity, or nudge the binary into a part of the parameter space where EKL can excite the binary eccentricity, both scenarios leading to a shorter GW merger time. However, before a binary can merge, it can become unbound due to cumulative encounters with other stars in the cluster \citep{BT}. This takes place on an ``evaporation" timescale of:
\begin{equation}\label{eq:evap}
    t_{ev} = \frac{\sqrt{3}\sigma(r)}{32\sqrt{\pi} G\rho(r) a_1\ln \Lambda}\frac{m_{1}+m_{2}}{m_{\rm m_b}} \ ,
\end{equation}
where $\rm ln~\Lambda = 15$ is the Coulomb algorithm, $m_b = 1~\msun$ is the average mass of background stars, and $\sigma(r)$ and $\rho(r)$ are given by Equations \ref{eq:sigma} and \ref{eq:rho} respectively. Note that our orbits are very eccentric, however, as shown by \citet{Rose+20}, the eccentricity may only change the evaporation timescale by a factor of a few. 

There are two pathways to merger, depending on whether EKL plays a significant part. In the case where $t_{\rm EKL} > t_{\rm GR,inner}$ (given by Equations \ref{eq:tEKL} and \ref{eq:tGR}), i.e. systems where GR effects dominate over EKL effects and the inner binary does not experience EKL-induced eccentricity excitations, we calculate the inner binary gravitation wave merger timescale following \citet{Peters64}:
\begin{equation}
t_{\rm GW} \sim  \frac{5}{265} \frac{c^5 a_1^4}{G^3 (m_1 + m_2) m_1 m_2}(1 - e^2)^{7/2}.
\end{equation}
If $t_{\rm GW} < t_{\rm Evap}$, we denote the system as a GW merger candidate.

In the case where $t_{\rm EKL} < t_{\rm GR,inner}$, we analytically estimate the maximal EKL-induced eccentricity, $e_{\rm 1,max}$ using the method detailed in \citet{Wen}, and estimate the EKL-induced GW merger time as:
\begin{equation}
 t_{\rm GW,EKL} \sim  \frac{5}{265} \frac{c^5 a_1^4}{G^3 (m_1 + m_2) m_1 m_2}(1 - e_{\rm 1, max}^2)^3,
\end{equation}
 \citep[e.g.][]{Liu+Lai18,Randall+Xianyu}. Using the above method, we find \GWmerger\ GW merger candidates per initial stellar binary (see Figure \ref{fig:AggregateStats} for break down by BH kick scenario). Note that this is likely an underestimate because the calculation of $e_{\rm 1,max}$ in \citet{Wen} uses the three-body Hamilton up to only the quadrupole order, and may miss octupole effects, which increases the merger rate \citep{Hoang+18}.  Using the same method we used to calculate the volume rate of EMRIs as in Section \ref{sec:EMRI}, we find a volume rate for GW mergers facilitated by natal kicks and EKL of 0.06-0.4 Gpc$^{-3}$ yr$^{-1}$.

\subsection{Binaries unbound from MBH}\label{sec:HVB}
Natal kicks can leave the inner binary bound while unbinding it from the MBH. In our simulations we find 0.007-0.01 binaries ejected from the MBH per initial stellar binary after one natal kick, and 0.005-0.006 after both natal kicks, making this the least common orbital configuration resulting from our simulations (see Figure \ref{fig:orb_config}). After both natal kicks, these unbound CO binaries are not observable. However, similarly to the ejected single objects from Section \ref{sec:unbound_single}, binaries ejected from the Galactic Center after $m_1$'s natal kick (these binaries contain one CO and one star), are observable. In the third column of Figure \ref{fig:HVS} we show the post-kick positions and velocities of these binaries after the first natal kick for the ``slow" BH kick case. Most of these binaries are ejected from the GC with velocities between 1000 - 10,000 km/s. These binaries are ejected from the Galactic Center between 4 and 45 Myr after the star formation episode, and $\sim 38$\% of these binaries contain a star that will not evolve to the CO phase until 100 Myr - 10 Gyr after the star formation episode, giving up to a few Gyr's window to observe them. 

Interestingly, $\sim 15$\% (17 \%) (22 \%) of the ejected binaries are X-ray binaries, for the no BH kicks (``slow" BH kicks) (``fast" BH kicks) scenario. For the ``fast" BH kicks scenario, 86\% (14 \%) of the ejected X-ray binaries have a NS (BH) primary. For both of the other BH kicks scenarios all the ejected X-ray binaries have a NS primary.

Of the ejected binaries that are not X-ray binaries,  $\sim 76\%$ (89\%)(96\%) have an NS primary, for the no BH kicks (``slow" BH kicks) (``fast" BH kicks) scenario. The rest have a BH primary.

We note that some previous theoretical works have predicted the existence of hypervelocity binaries \citep[e.g.][]{Fragione+17,Fragione+18,Wang+19}. Furthermore, a hypervelocity binary candidate has recently been observed \citep{Nemeth+16}.

\subsection{Binary Disruption by MBH}\label{sec:bin_disruption}

\begin{figure}
    \centering
    \includegraphics[width=\linewidth]{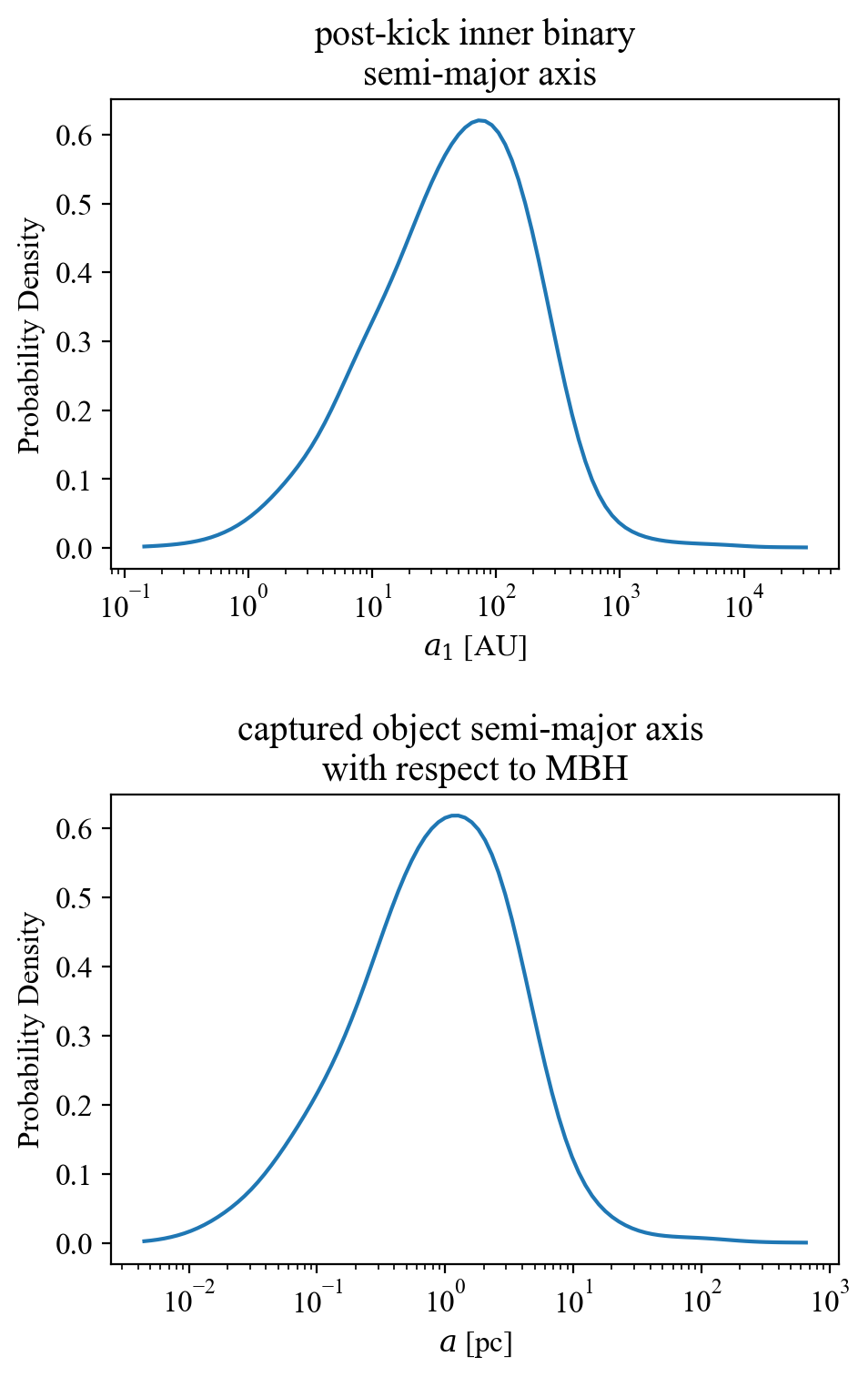}
    \caption{\textbf{The semi-major axis distribution of binaries disrupted by the MBH post-natal kick} (top panel), and the semi-major axis distribution of the binary component captured by the MBH (bottom panel).}
    \label{fig:bin_disr}
\end{figure}
In our Monte Carlo runs, we make sure that the inner binary birth distribution does not cross the Roche Limit of the MBH (Equation \ref{eq:RocheCrossing}). However, sometimes the natal kick will push the inner binary onto an orbit around the MBH that does  violate the condition given in Equation  (\ref{eq:RocheCrossing}). In this case the inner binary will be disrupted at the next pericenter passage with the MBH. In our simulations we find \BinDisr\ binaries disrupted by the MBH, per initial stellar binary. 

Typically, the heavier member of the inner binary is captured by (ejected from) the MBH for a bound (unbound) outer orbit, and vice versa for the lighter member \citep{Hills88,Yu+03,Sari+10,Kobayashi+12}. Previous works have shown that captured COs and stars from these binary disruption events can results in EMRIs, plunges, or TDEs \citep[e.g.][]{Sari+Fragione19,Miller+05}. We check whether this is the case for any of our binary disruptions. The inner binary is tidally disrupted by the MBH when it passes inside the tidal radius:
\begin{equation}
r_t = a_{1} \Big(\frac{m_{\rm MBH}}{m_1 + m_2}\Big)^{1/3}.
\end{equation}
The newly captured object's orbit has a pericenter equal to this tidal radius. Its new semi-major axis is given by:
\begin{equation}
    a_{\rm capt} \sim a_1 \Big(\frac{m_{\rm MBH}}{m_1 + m_2}\Big)^{2/3}
\end{equation}
We show the probability density of the post-kick inner binary semi-major axis, and the the probability density of the captured objects in Figure \ref{fig:bin_disr}. Because the binaries disrupted by the MBH are wide by definition (their distribution peaks at $\sim 100 {\rm AU}$), the captured objects have relatively large semi-major axes with respect to the MBH (their distribution peaks at $\sim 1$ pc). As a result, we find that none of the captured objects in our simulations result in EMRIs, plunges, and TDEs. However, they might have an effect on the stellar cusp in the inner parts of the Galactic Center \citep[e.g.][]{Fragione+Sari18}. 

The other binary member is typically ejected from the MBH at high velocities \citep[e.g.][]{Hills88}, and may add to the number of hypervelocity stars in our simulations. However, this out of the scope of this paper.

\section{Discussion}

We have conducted an analysis of the consequences of natal kicks in massive binaries at the Galactic Center. When a star undergoes supernova to become  BH or NS it is expected to receive a large natal kick \citep[e.g.][]{Hansen+Phinney97,Lorimer+Bailes97,Cordes+Chernoff98,Fryer+99,Hobbs+04,Hobbs+05,Beniamini+Piran16}. 

We run three large sets of Monte-Carlo simulations ($100,000$ for each set), calculating the dynamical outcome after supernova kick. The three sets correspond to three different kick distributions for BH progenitors: fast kicks (for which the kicks are similar to NS kicks), slow kicks (for which the kicks are normalized by the BH mass), and no kicks. The application of the kicks are done by simple vector analysis. The consist form of the equations are given in \citet{Lu+19}.

Generally, after each kick there are two main outcomes, where the inner binary becomes unbound or stays bound. The second kick may also unbind a binary in the latter case. After two kicks took place we find that most of the binaries were disrupted, e.g., between $94.5\% - 80.1\%$, for the fast to no kicks models respectively.  See Table \ref{tab:innerbinary} for the full details. However, even in the cases of inner-binary disruption, the majority of the systems remain bound to the MBH (see Table \ref{tab:singles}). These consequences yield the following predictions and observational signatures: 
\begin{itemize}
    \item \textbf{ Natal kicks create X-ray binaries}, at a rate of $(6.3-9.9)\times 10^{-3}$ per initial stellar binary. In general we find that faster kicks create more X-ray binaries. In particular, we find that no BH X-ray binaries are created if we assume no BH kicks. This suggests that detecting BH X-ray binaries in the Galactic Center implies some degree of BH kicks.  However, other formation channels can create BH X-ray binaries in the absence of kicks \citep[e.g.,][]{Naoz+16,Stephan+19}.
    \item \textbf{ Hypervelocity stars and COs are a common outcome.} We find that after the two kicks have taken place, a fairly significant fraction of the systems became unbound to the MBH (slightly more than $30\%$ of NSs, and between $32.3\%-1.7\%$ of BHs for fast to no kicks, respectively). See Table \ref{tab:singles}, for details. After the first kick, a number of stars are ejected from the inner 0.1 pc with velocities between 1000 - 10,000 km/s. Some of these stars may be observed as hypervelocity stars outside of the Galactic Center. 
    \item \textbf{Hypervelocity binaries} are created in addition to the ejected COs and stars, about 20\% of which are X-ray binaries. These binaries are ejected from the inner 0.1 pc after the first kick with velocities between 1000 - 10,000 km/s. In particular, after the first natal kick, we find $\sim (7.4 - 9.5)\times 10^{-3}$ binaries (CO + star) ejected from the inner 0.1 pc. These may be observed as hypervelocity binaries  and hypervelocity x-ray binaries outside of the Galactic Center.
    \item \textbf{Kicks result in a steeper distribution of the bound single NSs and BHs about $\sim 0.04$~pc from the MBH} compared to the initial distribution. The progenitors began on \citet{Bahcall+76} density profile, however, after the two kicks both the NS and the BHs (with fast kicks) result in a steeper distribution inwards to $0.04$~pc. At this distance from the MBH,  orbital velocity around the MBH is comparable to the average kick velocity of $\sim 200$~km~s$^{-1}$. Beyond this  radius, we find that the distribution of the NS and BHs (with fast kicks) have a {\it shallower} density distribution compared to their birth distribution. BHs with no kicks and WD (without kicks) follow the initial density distribution. This behavior is depicted in Figure \ref{fig:BoundSingles}.
    \item {\bf Kicks result in a slightly shallower binary density distribution about MBH.} Unlike the steeper segregation of singles toward the MBH, the bound inner binary post kicks result in a somewhat shallower density distribution, with a long tail of systems reaching large densities. Additionally, the inner binary semi-major  axis distribution after two natal kicks has a greater spread and peaks at a larger value, relative to the initial distribution.
    \item {\bf Exotic events} from our simulation include binary GW mergers, EMRIs, direct plunges, and TDEs. Of these, the most common are binary mergers, which occur in our simulations at a rate of $\sim$ \GWmerger\ per initial binary. EMRIs, direct plunges, and TDEs are much rarer events in our simulations. For each of these, we find a few times $10^{-5}$ events per initial binary. However, we note that our estimate of the EMRI rate is likely a conservative one due to assumptions about the inner edge of binary distribution around the MBH, see Section \ref{sec:EMRI} for further discussion.
\end{itemize}

In summary, we showed that natal kick have a  significant effect on the density distribution of COs, and of binaries in the galactic center. For example, it produces a steeper cusp of NSs about the MBH. Additionally, natal kicks naturally give rise to hypervelocity stars, hypervelocity binaries and X-ray binaries. Interestingly, we found that X-ray binaries can serve as a discriminator between BH kicks distribution models. Lastly, natal kicks can also result in a non-negligible rate of exotic events such as TDEs, LIGO, and LISA sources. 

\acknowledgments

BMH and SN acknowledge the partial support from NASA No. 80NSSC19K0321,  ATP-80NSSC20K0505 and  partial support from the NSF through grant No. AST-1739160. BMH thanks the University of California Office of the President Dissertation Year Fellowship.  SN thanks Howard and Astrid Preston for their generous support.

\bibliography{Binary}
\end{document}